\documentstyle[a4,12pt,epsfig,graphicx]{article} 
\def\beq{\begin{equation}}  
\def\eeq{\end{equation}}  
\def\bea{\begin{eqnarray}}   
\def\eea{\end{eqnarray}}   
\def\bq{\begin{quote}}   
\def\eq{\end{quote}}   
\def\bi{\begin{itemize}}   
\def\ei{\end{itemize}}   
\def\beqa{\begin{eqnarray}}   
\def\eeqa{\end{eqnarray}}   
\def\be{\begin{enumerate}}   
\def\ee{\end{enumerate}}   
\def\beq{\begin{equation}}   
\def\eeq{\end{equation}}   
\def\bi{\begin{itemize}} 
\def\ei{\end{itemize}}

\def\cq{{\cal Q}}

\def\r2{\sqrt{2}}   

\def\bi{\begin{itemize}}   
\def\ei{\end{itemize}}

\def\ca{{\cal A}}

\begin{document}
\titlepage
\begin{flushright}
hep-th/0303167 \\
\end{flushright}
\vskip 1cm
\begin{center}
{\large \bf Cosmological constant and kinetic supersymmetry 
breakdown on a moving brane}
\end{center}
\vspace*{5mm} \vspace*{1cm}  
%\end{center} 
\vspace*{5mm} \noindent 
\vskip 1cm 
\centerline{\bf Philippe Brax${}^{1}$, Adam Falkowski${}^{2}$ and Zygmunt Lalak${}^{2}$} 
\vskip 1cm 
%\centerline{\em ${}^{1}$Theory Division, CERN} 
%\centerline{\em CH-1211 Geneva 23, Switzerland} 
\vskip 0.3cm 
\centerline{\em ${}^{1}$ Service de Physique Th\'eorique} 
\centerline{\em CEA-Saclay F-91191 Gif/Yvette, France} 
\vskip 0.3cm 
\centerline{\em ${}^{2}$Institute of Theoretical Physics} 
\centerline{\em University of Warsaw, Poland} 
\vskip 2cm 
\begin{center}
{\bf Abstract}
\end{center}

We consider cosmological solutions in 5d locally supersymmetric
theories including  boundary actions, with either opposite tension branes for 
identical brane chiralities or equal tension branes  for  flipped brane chiralities. We analyse the occurrence of supersymmetry breakdown in both situations. 
We find that supersymmetry as seen by a brane observer is broken due to the 
motion of the brane in the bulk. 
When the brane energy-momentum tensor is  dominated by the brane tension, the 4d  vacuum energy %cosmological constant 
on the observable  brane is positive and proportional to the inverse 
square of the brane local time. We find that the mass splitting within supersymmetric
multiplets living on the brane 
is of the order of the inverse of the brane local time, 
examplifying  the tight relation between the vacuum 
energy scale and the supersymmetry breaking scale.

\newpage
\section{Introduction}
One possible interpretation of the cosmological evolution of the Universe is that it is due to the  motion of a brane embedded into a higher-dimensional ambient space. 
The brane moves according to the equations of motion, and its dynamics as well as the physics of the localized sector respond to local conditions encountered along its trajectory. 
In this paper we would like to study the dynamics of supersymmetric  bulk and branes, and their mutual coupling, in terms of a local action. In such a case the cosmological evolution may be related to the supersymmetry breakdown as seen on the brane. The aim  of the present work is to analyse  this    relationship  in an explicit way. 
To this end we use the brane-bulk supergravities constructed in \cite{Altendorfer:2000rr}, \cite{Falkowski:2000er},\cite{kallosh},\cite{Brax:2001xf},\cite{Brax:2002vs},\cite{Lalak:2002kx},\, with matter sectors localized on the  
branes and coupled supersymmetrically to the 5d supergravity with $AdS_5$ bulk. 
We find time-dependent solutions of the equations of motion that break 4d supersymmetry everywhere, in the bulk and on the branes. The observer on the visible brane finds a positive effective vacuum energy which changes as the inverse square of the local time. The scale of this `quintessential' 
vacuum energy is the same as the scale of susy breaking on the brane, hence asymptotically the observable brane becomes flat and supersymmetric. 
Of course, this is only  true  when one neglects the contribution of the matter sector. 

We investigate  the effective 4d supergravity that describes the local dynamics of the brane-bulk system in the vicinity of the brane. 
It turns out that the transmission of supersymmetry breakdown to the observable sector occurs via the radion multiplet (i.e. the T modulus). 
To make the model suitable to describe 
very late stages of the evolution of the Universe one needs to enhance the local scale of  brane supersymmetry breakdown with respect to the scale of the cosmological constant. 
The correspondence  of the 5d picture to 4d sugra with $T$-modulus suggests that the relevant enhancement mechanism would be equivalent to the tuning of the (super-)potential of the $T$-modulus and related to its stabilization. 

\section{Supergravity with Boundary Branes}

We discuss the cosmological evolution and the supersymmetry breaking of   5d N=2 supergravity with a negative cosmological constant. The fifth dimension is an interval of  finite length ending with   two branes located at $x^5 = r_0$ and $x^5=r_1$. These branes have their own cosmological constants (brane tensions) and are coupled to the 5d bulk in a way that preserves N=1 supersymmetry \cite{Altendorfer:2000rr},\cite{gp},\cite{Falkowski:2000er}. 
 We concentrate on a special subclass of these models, in which   the brane tensions are correlated with the bulk cosmological constant as in the Randall-Sundrum model \cite{rs1}.        

Let us first  recall the basic features of 5d N=2 gauged supergravity \cite{agata}. The gravity multiplet $(e_\alpha^m, \psi_\alpha^A, \ca_\alpha)$, $A=1,2$, consists of  the vielbein (metric), a pair of symplectic Majorana gravitini, and a vector field called the graviphoton. There is a global SU(2) R-symmetry which rotates the two supercharges into each other. Making use of the graviphoton we can gauge a U(1) subgroup of the R-symmetry group. What is important here is that the gauging implies the  presence of a negative cosmological constant in the bulk, $M^{-3} S_5 = \int d^5 x ( {1 \over 2}R +  6 k^2 + \dots)$ where $\kappa_5^2=M^{-3}$ is the 5d gravitational constant. Also, the  transformation law of the gravitini is altered by the gauging. The gravitino variation reads    
\beq 
\label{susygrav} 
\delta \psi_\alpha^A = D_\alpha \epsilon^A -  {1 \over 2} i g k \gamma_\alpha  \epsilon^A + \dots 
  \, , 
\eeq        
where $g$ is the $U(1)_R$ gauge coupling and the dots stand for terms involving graviphoton, which are not relevant for our discussion.

It is possible  to couple  branes in a supersymmetric way  even when the brane actions contain only the brane tension terms  \cite{gp},\cite{Falkowski:2000er}. To see this,  notice  that one can reformulate the theory on the interval (dowstairs picture) as a theory defined on a circle (upstairs picture). The values of  a field at the two sides of a brane are related by a $Z_2$ symmetry. We assume that, for the brane at $x_5=r_0$, the $Z_2$ symmetry acts on the gravitino as:  
\bea 
\label{z2} 
\psi^A_\mu(r_0-x^5)= \gamma_5 (\sigma^3)^A_B \psi_\mu^B(r_0+x^5) 
& \psi^A_5(r_0-x^5)= -\gamma_5 (\sigma^3)^A_B \psi_5^B(r_0+x^5) \, .
\eea 
Similarly, the $Z_2'$ symmetry relates the values of the gravitini  close to the $x^5 =r_1$ brane as       
\bea 
\label{z2f} 
\psi^A_\mu(r_1-x^5)= \alpha \gamma_5 (\sigma^3)^A_B \psi_\mu^B(r_1+x^5) &
 \psi^A_5(r_1-x^5)= -\alpha\gamma_5 \cq^A_B \psi_5^B(r_1+x^5). 
\eea   
We assume $\alpha$ can take values $+1$ or $-1$. For $\alpha=-1$ each brane respects  a different of the two bulk supersymmetries, so globally supersymmetry is entirely  broken.  
 
The point is that the gravitino variation has to have a jump at the branes, to be consistent with the $Z_2$ symmetry. This jump is a source of additional supersymmetric variation of the action, proportional to $\delta(x^5 - r_0)$ and $\delta(x^5 - r_1)$. These can be cancelled by a suitably chosen brane term representing a brane tension. For the case at hand the bosonic part of the Lagrangian that locally preserves $N=1$ supersymmetry is given by: 
\beq 
M^{-3} S= \int d^5 x \sqrt{-g_5} (\frac{1}{2}R + 6 k^2)-  6 \int d^5 x\sqrt{-g_4}k  (\delta(x^5-r_0) - \alpha \delta(x^5-r_1)) \, . 
\eeq 
For $\alpha=1$ the above is the Randall-Sundrum action \cite{rs2}. 
For $\alpha=-1$ the two brane tensions are equal. We call this possibility the flipped case.   

\section{Branes in motion}

We have described the Lagrangian coupling the $N=2$ supergravity background to boundary branes in a locally supersymmetric way.
We will now consider the vacua of such theories. In the bulk the vacua are determined by the equations of $N=2$ supergravity.
For the bosonic fields this corresponds to finding solutions of the Einstein equations with a negative bulk cosmological constant.
The five dimensional version of Birkhoff's theorem prescribes the most general solution in the bulk \cite{kraus}
\begin{equation}
ds^2= -f(r)dt^2 +\frac{dr^2}{f(r)}+r^2 d\Sigma_q^2,
\end{equation}
where the three-dimensional subspace spanned by the coordinates $\Sigma^i, i=1..\dots 3$ is respectively a spherical, flat or hyperbolic
space for $q=-1,0,1$. The function $f(r)$ is defined as
\begin{equation}
f(r)= k^2 r^2 +q -\frac{C}{r^2}.
\end{equation}
%Here we have introduced  $l=1/k$ where the bulk cosmological constant is $\Lambda=-12k^2$.
The parameter $C$ is related  to the mass of the black-hole located at 
the origin of coordinates.

The vacua are specified by the bulk metric and the motion of the two boundary branes.
This motion of a single brane follows from the Friedmann equation 
\cite{kraus}\cite{bowcock}
\begin{equation}
H^2= -\frac{q}{r^2}+\frac{C}{r^4}
\end{equation}
for $H=\dot r/ r$ where the dot denotes $d/d\tau $ and $\tau$ is the proper time on the brane such that the
induced metric on the brane reads
\begin{equation}
ds_B^2=-d\tau^2+ r^2(\tau)d\Sigma_q^2.
\end{equation}
The first term in the Friedmann equation is 
the usual curvature contribution while the second term is  the so-called dark radiation.
The solutions to the Friedmann equation are easily found.
For $q=0$ we obtain
\begin{equation}
r(\tau)=\sqrt{\pm 2(\tau-\tau_0)}.
\end{equation}
On the brane the geometry is the one of a radiation dominated FRW space-time.
The brane either goes to infinity or is driven towards the origin at $r=0$.
For $q=-1,1$ the solutions are very different:
\begin{equation}
r(\tau)= (\frac{C -(\tau-\tau_0)^2}{q})^{1/2}.
\end{equation}
For $q=1$ the brane cannot escape to infinity, it may reach  a maximal 
coordinates distance of $r_{max}=\sqrt C$ before bouncing 
back towards the singularity.
In the open case $q=-1$ the brane does not return to the singularity, 
and may reach infinity. Notice too that as $\tau$ goes to infinity, the motion of the brane becomes
identical with the case where $C=0$. The induced geometry on the brane corresponds to Milne space-time (see appendix
A for more details).

Let us now consider the motion of a two-brane system. We have to treat the 
unflipped and flipped cases separately.
Let us first concentrate on the unflipped case where the brane 
tensions are opposite.

The bulk parameters $C$ and $q$ are determined by the initial positions and velocities of the branes   
(the branes are specified by their position $r_0$ and $r_1$, and their velocities $\dot r_0$ and $\dot r_1$).
 The Friedmann equation implies that 
\begin{equation}
C= r_0^2r_1^2 \frac{\dot r_1^2- \dot r_0^2}{r_0^2-r_1^2}.
\end{equation}
As soon as the initial velocities of the branes are different, the black-hole 
mass
$C$ has to be non-zero. 
The initial velocities are constrained by 
\begin{equation}
q= \frac{r_1^2\dot r_1^2-r_0^2\dot r_0^2}{r_0^2-r_1^2}.
\end{equation}
Notice that $q=0$ and $C=0$ is only compatible with static branes.
In all other cases either $q$ or $C$ must be different from zero.

\section{Bulk  Supersymmetry Breakdown}

We have seen that the vacua are characterized in the bulk by two parameters,
the black-hole mass $C$ and the curvature $q$.
As soon as the branes move, one of these parameters 
is non-vanishing. We will now show that this implies that supersymmetry is broken by the motion of the branes. In this section  we show that in the background with non-zero $q$ or $C$ it is impossible to define Killing spinors.
\subsection{Killing Spinors}

Let us first define Gaussian normal coordinates $x_5$
\begin{equation}
x_5=\frac{l}{2}(\theta-\ln(\frac{r_+^2+r_-^2}{4l^2})),
\end{equation}
where
\begin{equation}
r^2=\frac{r_-^2+r_+^2}{2}\cosh\theta +\frac{r_+^2-r_-^2}{2},
\end{equation}  
such that
the metric becomes
\begin{equation}
ds^2= dx_5^2
-\frac{r_-^2+r_+^2}{4l^2}\frac{\sinh ^2 (\theta)}{\frac{r_-^2+r_+^2}{2}\cosh (\theta)+\frac{r_+^2-r_-^2}{2}}
dt^2 + (\frac{r_-^2+r_+^2}{2}\cosh (\theta)+\frac{r_+^2-r_-^2}{2})
d\Sigma_q^2.
\end{equation}
Let us identify this form of the metric with
$ds^2=dx_5^2-f^2(x_5)dt^2 +g^2(x_5) d\Sigma_q^2$.
We can now read  the spin connection
\begin{equation}
\omega_{t\hat t\hat 5}=-f',\ \omega_{k\hat i\hat j}=\tilde
\omega_{k\hat i\hat j},\ \omega_{k\hat i \hat 5}= g'\tilde e
_{\hat i k}
\end{equation}
where hatted indices are local frame indices, $'=d/dx_5$ and the
tilded symbols refer to the symmetric space of curvature $q$.

Let us now analyse the Killing spinors in the bulk. They satisfy
\begin{equation}
D_{a}\epsilon^A -\frac{k}{2}\gamma_{a}(\sigma^3)^A_B \epsilon^B=0
\end{equation}
This leads to
\begin{equation}
\partial_5\epsilon^A=\frac{k}{2}
\gamma_{5}(\sigma^3)^A_B \epsilon^B.
\end{equation}
Define now two spinors
\begin{equation}
(\epsilon^{\pm})^A=\pm \gamma_5 (\sigma^3)^A_B(\epsilon^{\pm})^B.
\end{equation}
This implies that
\begin{equation}
\epsilon^\pm(x_5,x^\mu)=e^{\pm \frac{k}{2}x_5}\epsilon^{\pm}(x^\mu).
\end{equation}
Now the other components of the Killing spinor equations lead to
\begin{equation}
\tilde
D_i(\epsilon^{\pm})^A+(\pm\frac{g'}{2g}-\frac{k}{2})\gamma_i
(\sigma^3)^A_B (\epsilon^\pm)^B=0,
\end{equation}
where $\tilde D_i$ is the covariant derivative on the symmetric
space of curvature $q$,
and
\begin{equation}
\partial_0 (\epsilon^{\pm})^A+(\pm\frac{f'}{2f}-\frac{k}{2})\gamma_0
(\sigma^3)^A_B (\epsilon^\pm)^B=0.
\end{equation}
The integrability conditions in the maximally symmetric curved space  
\begin{equation}
[\tilde D_i,\tilde D_j]=\tilde R_{i\hat aj \hat
b}\frac{\gamma^{\hat a\hat b}}{4}
\end{equation}
with
\begin{equation}
\tilde R_{abcd}=q(\tilde g_{ac}\tilde g_{bd}-\tilde g_{ad}\tilde g_{cd})
\end{equation}
give
\begin{equation}
\frac{q}{g^2}\gamma_{ij}\epsilon^\pm=(\frac{k}{2}\mp\frac{g'}{2g})^2
\gamma_{ij}\epsilon^{\pm}.
\end{equation}
This equation leads to two different possibilities.
If $q=0$ and $C=0$ then $g'/2g=k/2$ implying that $\epsilon^-=0$
and $\epsilon^+$ is left undetermined.
Coming back to the Killing equations we find that $\epsilon^+$ is
independent of $x^\mu$. This is the usual BPS situation
where $1/2$ of supersymmetry is preserved by the vacuum.
In all the other cases, i.e. as soon as the branes start moving
the only solution is $\epsilon^\pm=0$ leading to a complete
breakdown of supersymmetry.

Hence we conclude that the motion of the branes is responsible
for the breakdown of supersymmetry. This breakdown is spontaneous as
realized due to the lack of supersymmetry of the vacuum solution, 
despite the supersymmetric invariance of the action.
%We shall analyse the signatures of this spontaneous breakdown of local supersymmetry.

\subsection{KK-modes of gravitini}

Let us consider  the gravitino equation of motion in the bulk,
\begin{equation} \label{graeq}
\gamma^{\alpha\beta\delta}D_\beta\psi^A_\delta-\frac{3}{2}k
(\sigma^3)^A_B \gamma^{\alpha\delta}\psi_\delta^B=0.
\end{equation}
The last term comes from the gravitino mass term in the bulk.
We work in the gauge where $\psi_5=0$.
Let us focus on the $A=1$ component, the $A=2$ component being
equivalent due to the symplectic-Majorana condition imposed on five
dimensional spinors.
Let us decompose
\begin{equation}
(\psi^1_0)_L=\phi^+(x_5) (\psi_0^+)_L(x^\mu), (\psi^1_0)_R=\psi^-(x_5) (\psi_0^-)_R(x^\mu),
\end{equation}
and 
\begin{equation}
(\psi^1_i)_L=\frac{g}{f}\phi^+(x_5) (\psi_i^+)_L(x^\mu), (\psi^1_i)_R=\frac{g}{f}\phi^-(x_5) (\psi_i^-)_R(x^\mu),
\end{equation}
where $\psi^\pm_\mu$ are four dimensional Majorana spinors.
Notice the explicit breaking of Lorentz invariance when $f\ne g$.

We are going to solve the gravitino equation separating the variables
and decomposing the wave function into a product of the
five-dimensional and four-dimensional factors. We shall denote the separation constant by $m$, and call it a mass term, although strictly speaking we do not identify this parameter with the physcial mass. The precise correspondence between these two 
can be made explicit, e. g. in the case of $AdS_4$ geometry in 4d, see \cite{Lalak:2002kx}, but here we consider more complicated 4d geometries. Whenever we use the
term `four-dimensional mass' we have in mind  
mass parameters in the sense of a general four-dimensional supergravity 
Lagrangian. 
Hence, we take the massive  four-dimensional gravitini to satisfy
\begin{equation}
\tilde \gamma^{\rho\nu\mu}D^4_\nu\psi^\pm_\mu= m\tilde \gamma^{\rho\mu}\psi^{\mp}_\mu,
\end{equation}
where $\tilde \gamma_\mu$ is the Dirac matrix on the
four-dimensional space-time with spatial curvature $q$.
Notice the necessary exchange of chiralities in the Weyl basis.

The gravitino equation (\ref{graeq}) can be treated in two steps. First of all,
putting $\alpha=0$ leads to the evolution equations
\begin{equation}
-\partial_5\phi^++\frac{3}{2}k \phi^+ -\frac{g'}{g}\phi^+=\frac{m}{g}\phi^-
\end{equation}
and
\begin{equation}
\partial_5\phi^-+ \frac{3}{2}k \phi^- +\frac{g'}{g}\phi^-=\frac{m}{g}\phi^+
\end{equation}
where $\gamma_5$ acts as $-1$ on left-handed spinors
The $\alpha=k=1,2,3$ equation leads to new constraints on the spinors.
We find that
\begin{equation}
(\frac{1}{gf^2}-\frac{1}{g^2f})\tilde \gamma^{k0i}\partial_0(\psi^-_i)_R\phi^-+ \frac{3}{2}(\frac{g'}{f}-\frac{f'}{g})\phi^+\gamma^{ki}(\psi^+_i)_L
=0.
\end{equation}
This is consistent provided we impose the irreducibility constraint
\begin{equation}
\tilde \gamma^i \psi_i^-=0, \ \tilde\gamma^i\psi_i^+=0,
\end{equation}
In conclusion we have found that the spectrum of gravitini corresponds to massive gravitini
of the 4d hyperbolic space together with an irreducibility constraint on spinors
on the 3d hyperbolic space. 

We can now examine the nature of the spectrum for both  equal
and opposite tension cases.
We concentrate on the zero modes ($m=0$) which  are determined by
\begin{equation}
\phi^+= e^{3kx_5/2}g^{-1}\phi_0^+, \phi^{-}=e^{-3kx_5/2}g^{-1}\phi_0^-.
\end{equation}
Notice now that $g$ does not vanish on the whole $x_5$ axis.
In the opposite tension brane, one can choose $\phi^-_0=0$ and then 
one obtains a single massless gravitino in four dimensions. The
remaining chirality is the same as the chirality of the Killing spinor.
When the chiralities are flipped, this implies that $\phi_0^\pm=0$ therefore
all the massless gravitini are projected out. This is the
expected result for flipped supersymmetry breaking.

\section{Brane Supersymmetry Breaking}

Let us concentrate on opposite tension branes 
moving away to infinity. This is the `moving brane' version of the static 
Randall-Sundrum setting. Asymptotically  the branes are located at
\begin{equation}
x_5^{(2)}=t,\ x_5^{(1)}=t-R,
\end{equation}
where $R$ is the radion field. 
The far away metric becomes now 
\begin{equation}
ds_5^2=dx_5^2 + e^{2 k x_5}g_{\mu\nu}dx^\mu dx^\nu
\end{equation}
where the background metric $g_{\mu\nu}=\tilde g_{\mu\nu}$ is the metric on the
hyperbolic space 
\begin{equation}
ds_0^2=-dt^2 +l^2d\Sigma^2.
\end{equation}

We will now perform the dimensional reduction  and express the Lagrangian 
in terms of the metric $g_{\mu\nu}$.
% To have a better intuition about the present setup, one could imagine the two-brane Randall--Sundrum--type model with nonzero initial velocities of the branes. The results presented below imply an instability in the purely gravitational system of this type. 
Let us  start with the vacuum energy due to the bulk and brane contributions. 
The brane contributions comprise the brane action and the
Gibbons-Hawking terms, \cite{Lalak:2001fd}. It is straightforward to show that 
the contributions of bulk and brane cancel so that the `hard'   4d cosmological constant vanishes altogether.

The Einstein-Hilbert action from the bulk becomes the four-dimensional
gravitational action for $g_{\mu\nu}$ with a gravitational coupling 
\begin{equation}
\frac{1}{\kappa^2}=\frac{\int_{t-R}^{t} e^{2 k x_5}}{\kappa_5^2}
\end{equation}
implying that in the large time regime, the
gravitational coupling  becomes
\begin{equation}
\frac{1}{\kappa^2}=\frac{a^2(t)}{(\kappa_4)^2 } e^{-K/3},
\end{equation}
where $a(t) = e^{kt}$ and 
\begin{equation}
\frac{1}{(\kappa_4)^2}=
\frac{2}{ k \kappa_5^2}.
\end{equation}
We have identified 
\begin{equation}
2 k R =T+ \bar T 
\end{equation}
as the real part of a scalar field which belongs to a chiral
superfield $T$. Such a chiral multiplet contains also the fifth
component of the graviphoton, which plays the role of an axion, and
the fifth component of the gravitino,  called the radino. 
The K\"ahler potential of the radion superfield has been identified as 
\cite{Falkowski:2001sq}
\begin{equation} \label{kah}
K=-3\ln(1-e^{-(T+\bar T)}).
\end{equation}

As $x_5>>l$ the form of the metric simplifies and $f=g=e^{2x_5/l}$. In that case, there is an 
asymptotic restoration of supersymmetry implying that the massless gravitini 
coincide with the ones of the static supersymmetric Randall-Sundrum model. Moreover the Killing spinors
are now non-vanishing corresponding to a local, i.e. occurring in the vicinity of the moving brane, restoration
of supersymmetry
We can identify the  zero mode for the gravitini
\begin{equation}
\psi_\mu=a^{1/2}(x_5)\hat \psi_\mu (x)
\end{equation}
where $\hat\psi$ is a four dimensional massless Majorana spinor.
Similarly one can solve for the Killing spinor 
with the result
\begin{equation}
\epsilon=a^{1/2}(x_5)\hat \epsilon (x),
\end{equation}
reflecting the fact that locally only 
one combination of supercharges is preserved in the bulk/brane system.
We also include the contribution coming from
\begin{equation}
\psi_5(x)= a^{-1/2}(x_5)\hat\psi_5(x).
\end{equation}
Denoting by $\hat D_\mu$ the covariant derivative in four
dimensions with respect to $g_{\mu\nu}$, we identify the low energy  Lagrangian
\begin{equation}
\int d^4x \sqrt{-g} \frac{1}{2\kappa^2}(R(g)+\overline{\hat \psi}^\mu\hat
\gamma_{\mu\nu\rho}\hat D^{\nu}\hat\psi^{\rho}+\overline{\hat \psi^\mu}\hat
\gamma_{\mu\nu}\gamma^5\hat D^{\mu} \hat\psi_5 +\overline{\hat \psi^5}\gamma_5\hat
\gamma_{\mu\nu}\hat D^{\mu} \hat\psi^\nu) 
\end{equation}
where $\hat \gamma^\mu$ are the 4d gamma matrices with respect to $g_{\mu\nu}$.
To remove the kinetic mixing between $\hat \psi_5$ and $\hat \psi_\mu$ we
redefine the fields
\begin{equation}
\hat\psi_\mu\to\hat\psi_\mu -\frac{1}{2}\hat\gamma_\mu\gamma^5
\hat\psi_5,
\end{equation}
which leads to the action
\begin{equation}
\int d^4x \sqrt{-g} \frac{1}{2\kappa^2}\sqrt{-g}(R(g)+\overline{\hat \psi}^\mu\hat
\gamma_{\mu\nu\rho}\hat D^{\nu}\hat\psi^{\rho}+\frac{3}{2}\overline{\hat \psi_5}\hat
\gamma_{\mu}\hat D^{\nu}\hat\psi_5  ). 
\end{equation}
 
Let us analyse the supersymmetry variations of the gravitini.
We find  that 
\begin{equation}
\delta_{\hat \epsilon}\hat \psi_\mu\supset \hat D_\mu  
\hat \epsilon+ \frac{k}{2}g^{0}_{\mu}\hat \epsilon.
\end{equation}
i.e. $\langle \delta_{\hat \epsilon}\hat \psi_\mu \rangle \neq 0$.
The first contribution is the usual 4d supergravity variation, the other
one  comes from  the time
dependence of the normalization factor of gravitini on the brane. 
This second term is the signal
of supersymmetry breaking. This is yet another example of 
supersymmetry breakdown due to the boundary conditions in the brane system,
this time along the time direction while in the Scherk-Schwarz case the 
new direction is space-like. We shall call the present mechanism a {\em 
kinetic} supersymmetry breaking. Of course, the kinetic breaking is 4d Lorentz non-covariant, hence it shall manifest itself somewhat differently than the usual breaking through the F-terms and D-terms.    
One would like to obtain the local interpretation of this susy breakdown given by a local observer bound to the brane. Locally, near the brane 
(here we choose the positive tension brane to be the observable one) it should be possible to understand the situation in terms of a local supergravity
defined in the vicinity of the brane. 

A crucial sign  of supersymmetry breaking is the presence of a
time-dependent  cosmological constant in the Einstein frame 
defined by $g^E_{\mu\nu}=a^2(t) e^{-K/3} g_{\mu\nu}$.
Using
\begin{equation}
\int d^4 x a^2(t) e^{-K(T)/3}\sqrt{-g} R(g)\supset \int d^4x
\sqrt{-g_E}(R(g_E)-6g_E^{\mu\nu}\partial_\mu\ln a\partial_{\nu}\ln a),
\label{pot}
\end{equation}
we find that in the Einstein frame
the initial time dependence of the gravitational constant becomes a
time dependent vacuum energy 
\begin{equation} \label{ccc}
\hat \Lambda =\frac{3}{\kappa_4^2 }(\frac{d \ln a}{d\tau})^2
\end{equation}
where $\tau$ is the brane local time\footnote{We reserve the label $\tau$
for the brane local time, while $t$ denotes the coordinate time.}.
Using $a(t)=\tau/l$ this leads to
\begin{equation}
\hat \Lambda =\frac{3}{(\kappa_4)^2 \tau^2}.
\end{equation}
This is a positive time dependent cosmological constant which decays as $\tau\to\infty$ corresponding to the
asymptotic restoration of supersymmetry. 

One can define canonically normalized gravitini and
radino in the Einstein frame
\begin{equation}
\tilde \psi _\mu (x)= a^{1/2}(t)e^{-K/12}\hat\psi_{\mu}(x),\
\tilde \psi _5 (x)= \sqrt{\frac{3}{2}}a^{-1/2}(t)e^{K/12}\hat\psi_{5}(x).
\end{equation}
Now in the Einstein frame the gravitino variation becomes
\begin{equation} \label{basiceq}
\delta_{\tilde  \epsilon}\tilde  \psi_\mu\supset  D^E_\mu  \tilde  \epsilon-
\frac{e^{K/6}}{8la(t)}\left [ \gamma^{\bar 0}, \hat \gamma^E_\mu\right ],
\end{equation}
where we have defined $\hat \epsilon= a^{-1/2}(t)e^{K/12}\tilde \epsilon$
and $D^E$ is the covariant derivative with respect to the Einstein metric.
The matrix $\gamma^{\bar 0}$ is the local frame gamma matrix.
We have used the fact that the induced metric leads to an extra
spin connection
\begin{equation}
\omega_{\bar\mu\bar0\nu}\supset \frac{e^{K/6}}{la(t)} e^E_{\bar\mu\nu}
\end{equation}
where $e^E_{\bar \mu \nu }$ is the veilbein of the metric
$g^E_{\mu\nu}$. Now $<\delta_{\tilde\epsilon}\tilde\psi_\mu>\ne 0$ due to the generator  associated
with a Lorentz boost.

The variation of the gravitino (the one propagating on the brane) in the 
effective 4d theory is 
\begin{equation}
\delta _{\tilde{\epsilon}} \psi_\mu = 2 D_\mu \epsilon + i e^{K/2} W \gamma_\mu \epsilon + ... \, ,
\end{equation} 
where $D_\mu = \partial_\mu + 1/4 \omega_{\mu}^{mn}[\gamma_m,\gamma_n]$. 
Comparing this to (\ref{basiceq}) we can see that the Lorentz structure of 
the variation induced by the brane motion is not that of the term associated with the expectation value of the superpotential $W$, but rather should be seen
as a time-dependent background connection.

Usually, in spontaneously broken supergravity a positive cosmological
constant is
associated with non-vanishing $F$ terms. Here the subtlety lies in the fact, 
that the vacuum solution we have at our disposal is time-dependent, 
and it is determined in the full 
theory containing both branes. It is not obvious that 
for instance in the flipped (twisted) case of \cite{Brax:2001xf} the four-dimensional description of the full system in terms of the standard 4d supergravity is possible. 
On the other hand, for an observer localized on the travelling brane, who has 
at his disposal local supercharges operating on the brane, it should be possible to describe the local dynamics in terms of sugra Lagrangian containing degrees of freedom that are allowed to fluctuate on the brane. We find such an approximate  
description possible, but the price to pay is  the explicit dependence on  time of the supersymmetry variations. In the case discussed in detail 
in this section it is natural to keep in the effective brane theory the 
degree of freedom corresponding to the relative motion of the branes, the radion field. The overall motion of the system with respect to the bulk 
is encoded in the explicit time dependence of the variations. 

Let us analyse first the type of geometry seen by a low energy observer
on the brane.
Taking the ansatz
\begin{equation}
g^E_{\mu\nu}=-d\tau^2 +a^2(\tau)dx^i dx_i
\end{equation}
where spatial sections are flat now.
The Friedmann equation reads
\begin{equation}
H^2\equiv \frac{\kappa_4^2}{3}\hat \Lambda =\frac{1}{\tau^2}
\end{equation}
implying that $a(\tau)=\tau$. 
Notice that the Hubble rate as seen from the effective action coincides with Hubble rate
obtained from the induced metric on the brane.

Let us now consider the coupling to supersymmetric matter on the brane. It is crucial to notice that matter
couples to the induced metric $g^B_{\mu\nu}=a^2(t)g_{\mu\nu}$ for far away branes.
The kinetic terms are therefore
\begin{equation}
S_{kin}=\int d^4x a^2(t) \sqrt {-g} g^{\mu\nu}\partial_\mu \Phi \partial_\nu \bar \Phi.
\end{equation}
One can deduce the complete K\"ahler potential including the radion superfield
$T$ and chiral matter $\Phi$ on the brane \cite{Falkowski:2001sq}
\begin{equation} \label{kpotmat}
K(T, \Phi) = -3 \log( 1 - e^{-(T + \bar{T})} - |\Phi|^2) .
\end{equation}
%Notice that it is not of the sequestered form.

Let us now go to the Jordan frame where matter on the brane couples
directly to the induced metric $g^B_{\mu\nu}=a^2(t)g_{\mu\nu}$
The action contains a term
\begin{equation}
\int d^4 x a^2(t) e^{-K(T,\Phi)/3}\sqrt{-g} R(g)\supset \int d^4x
\sqrt{-g_B}e^{-K(T,\Phi)/3}(R(g_B)-6g_B^{\mu\nu}\partial_\mu\ln
a\partial_{\nu}\ln a),
\end{equation}
Using the explicit form of the K\"ahler potential and time dependence of 
$\ln a$, this leads to a soft breaking mass term for the chiral
superfields
living on the brane
\begin{equation}
\Delta m^2 = \frac{3}{\tau^2}, 
\end{equation} 
and this is equal, up to the powers of the 4d Planck scale  to the cosmologicalconstant. 
Notice that the splitting within the matter multiplets is also related to the Hubble rate on the brane
\begin{equation}
\Delta m^2= 3H^2
\label{Fried}
\end{equation}
showing the intrinsic link between the kinetic breaking of supersymmetry and the cosmological evolution of the brane.
As expected the mass splitting dies off as the brane recedes away towards infinity. 

One may give a simple explanation for the appearance of a direct link between the mass splitting and the Hubble rate.
This relationship  comes  from the kinetic term for the modulus $T$, the only bulk modulus that is left in the effective local sugra. The necessary ingredient 
is the presence of the K\"ahler potential mixing between $T$ and chiral multiplets $\Phi$. This mixing does exist indeed, as seen in (\ref{kpotmat}),
and results in 
\begin{equation}
L_{kin} = g_{T \bar{T}} |\frac{\partial T}{\partial \tau}|^2 = 
3 |\frac{\partial T}{\partial \tau}|^2 e^{-(T + \bar{T})} (1+|\Phi|^2) + ... \, ,
\end{equation}
hence the (unnormalized) mass terms are 
\begin{equation}
m^{2}_{\Phi} = 
3 |\frac{\partial T}{\partial \tau}|^2 e^{-(T + \bar{T})}.
\end{equation}
In this frame the mass terms are induced by the nontrivial time evolution of the modulus $T$ when we identify the kinetic energy of the modulus $T$ with the Hubble rate as given by the five-dimensional dynamics, i.e. 
the Friedmann equation 
\begin{equation}
3 H^2 =  g_{T \bar{T}} |\frac{\partial T}{\partial \tau}|^2= 3 |\frac{\partial T}{\partial \tau}|^2 e^{-(T + \bar{T})}, .
\end{equation}
This again leads to a mass splitting proportional to the cosmological scale 
$H$ (\ref{Fried}).

The result obtained in this section has also another  nice interpretation in terms of temperature supersymmetry breaking on the brane.
Focusing on one of the moving brane, it is easy to see that the asymptotic geometry is of the Milne type
(see  appendix A), i.e. a flat brane. The motion of the brane in the bulk implies
that an observer on the brane coupled to the bulk vacuum detects a thermal spectrum of temperature \cite{das}
\begin{equation}
T_B=\frac{1}{2\pi \tau}.
\end{equation}
Now in flat Minkowski space, a globally supersymmetric theory such as the MSSM is broken 
by temperature effects. The mass splitting of the supersymmetric multiplet is of order $O(T_B)$.
%with the same time variation as obtained from the effective action.
This is not surprising, since also the temperature supersymmetry breakdown can be understood as the result of nontrivial boundary conditions imposed on 
bosons and fermions along the Euclidean time direction.

\section{Flipped Supersymmetry Breaking}

\subsection{Branes beyond the Horizon}

Let us now come back to the flipped case. The two branes have
positive tensions preventing the embedding of the two branes on the
same side of the Einstein-Rosen bridge at $U=V=0$ (see appendix A for details on the global structure of space-time).
However, one can embed the branes on both sides of the Einstein-Rosen
bridge. In that case the two branes are separated by a wormhole.
The initial condition problem can be analysed and leads to the
same conclusions as for the opposite tension case.
Here the two branes are in causally disconnected regions.

For far away branes we focus on the AdS wormhole $q=-1,\ C=0$.  
The global structure of space-time consists of 
two copies of the same $r\ge l$ space glued at $r=l$ by a wormhole whose topology is the one of $\Sigma_{-1}$. 
To see that one can use the isotropic coordinates defined by
\begin{equation}
r=l\hbox{cosh}(\ln\tan (\frac{1}{2}\ln \frac{\rho}{l}+\frac{\pi}{4}))
\end{equation}
such that the metric becomes
\begin{equation}
ds^2=-\hbox{sinh}^2(\ln \tan (\frac{1}{2}\ln \frac{\rho}{l}+\frac{\pi}{4}))dt^2+\frac{l^2\hbox{cosh}^2(\ln \tan (\frac{1}{2}\ln \frac{\rho}{l}+\frac{\pi}{4}))}
{\rho^2}(d\rho^2+\rho^2d\Sigma^2)
\end{equation}
Notice that this parametrization realizes explicitly $r\ge l$. Moreover the relation between
$r$ and $\rho$ is two to one, i.e. the coordinate $\rho\in [0,\infty)$ covers the line
$r\ge l$ twice. This is the origin of the two sides of the horizon.
A more precise description can be obtained by noticing that the two sides of the horizon are
in fact  exchanged by the isometry
\begin{equation}
\rho\to\tilde \rho= \frac{l^2}{\rho}
\end{equation}
The fixed point of this transformation
\begin{equation}
\rho_c=l
\end{equation}
is precisely located at the horizon.

On both sides of the horizon, the branes follow motions identical 
to the one obtained in the opposite tension case. 
The main difference comes from the fact that the horizon is now
visible for both branes. This leads to a greatly modified behaviour of gravity
due to the strong Lorentz violation in the region close to the horizon. 
However, the contribution to the four-dimensional Planck scale coming from 
the near-horizon region is finite and necessarily much smaller (see the next 
subsection) than the contributions from the near-brane regions, because the 
warp factor grows towards the branes. Hence   
the effective supergravity/matter model on the positive tension brane in the 
asymptotic far-away region of the space-time 
is the same as in the `moving' RS case discussed earlier.

\subsection{Gravity and Lorentz Violation}

We will now consider the graviton  in the AdS wormhole with $q=-1, \ C=0$ corresponding to a
moving brane receding away towards infinity.
We work in the trace-less transverse gauge $h=0$ and $D_ah^{ab}=0$.
The zero modes can be written in the form
\begin{equation}
h_{ij}=e_i^ue_j^v h_{uv}
\end{equation}
where $(i,j)$ are indices of the symmetric space of curvature
$q$. The graviton depends on the vielbein $e_i^u$ on the
symmetric space . The matrix $h_{uv}$ is symmetric and traceless,
corresponding to five degrees of freedom. Gauge invariance allows
to fix three degrees of freedom, leaving two independent polarizations.
Zero modes correspond to time-independent matrices $h_{uv}$.

The  graviton equation is easily obtained by noticing that the
AdS wormhole ($q=-1,\ C=0$) is locally isometric to AdS space-time and therefore \cite{chris}
\begin{equation}
D_aD^a h_{ij}+2k^2 h_{ij}=0
\end{equation}
Using the connections
\begin{equation}
\Gamma^0_{05}=\frac{f'}{f},\ \Gamma_{00}^5=\frac{f'}{f}
\end{equation}
and
\begin{equation}
\Gamma^{i}_{j5}=\frac{g'}{g}\delta^i_j,\ \Gamma_{ij}^5=-\frac{g'}{g}g_{ij}
\end{equation}
we find that the graviton equation reads
\begin{equation}
h_{ij}'' +(\frac{f'}{f}-\frac{g'}{g})h_{ij}' -2[(\frac{g'}{g})'+ \frac{g'}{g}(2\frac{g'}{g}+\frac{f'}{f})-k^2]h_{ij}=0
\end{equation}
Close to the horizon we have $\frac{f'}{f}\approx  \frac{1}{x_5},\ \frac{g'}{g}\approx 0$
implying that
\begin{equation}
h_{ij}''+\frac{1}{x_5}h_{ij}'+2k^2h_{ij}=0
\end{equation}
Putting $h_{ij}= x_5^{-1/2}H_{ij}$ we obtain that
\begin{equation}
H''_{ij}+\frac{1}{4x_5^2}H_{ij}=0
\end{equation}
%This is the universal behaviour of gravitons noticed in a
%previous paper.
and  close to the horizon
\begin{equation}
h_{ij}= h^0_{ij}+ \ln \vert x_5\vert  h^{1}_{ij}
\end{equation}
where $h_{ij}^{0,1}$ are constant tensors in terms of $x_5$.

Let us now evaluate the kinetic terms of the graviton
\begin{equation}
\frac{1}{2\kappa_5^2}\int dx_5 d^4 x fg^{-3} \phi^2 (\partial
\tilde h)^2
\end{equation}
where $h_{ij}=\phi(x_5) \tilde h_{ij}$ and the contractions are
made in flat space. Integrating out the extra dimension between branes 
yields an effective Planck scale
\begin{equation}
\frac{1}{2\kappa_4^2}=\frac{1}{2\kappa_5^2}\int dx_5 fg^{-3} \phi^2.
\end{equation}
This is well defined everywhere except at the horizon where the integral
behaves like
\begin{equation}
\int dx_5 x_5 \ln ^2(x_5)<\infty.
\end{equation}
Hence the effective Planck mass is finite.

Notice that close to the horizon, Lorentz invariance is broken
implying that one cannot write a gravity theory in the usual
Einstein Hilbert form.
It is only at spatial infinity where $f=g$ that Lorentz
invariance is restored, allowing to consider more familiar
looking effective actions.

\section{Summary}

In this paper we have discussed the physics on a non-static brane of locally supersymmetric five-dimensional supergravity which 
moves with respect to the bulk according to the classical equations of motion. 
We have found that in such a case 
the cosmological evolution is related to the dynamics of the supersymmetry breakdown seen on the brane. 
Within the  brane-bulk supergravity of \cite{Falkowski:2000er},\cite{Brax:2001xf}
there exist time-dependent solutions that break 4d supersymmetry everywhere, in the bulk and on the branes. The observer on the visible brane finds a positive effective vacuum energy which changes as the inverse square of the local time. Having obtained a decaying cosmological constant, one may wonder
if this is not the realization of quintessence. In fact one finds that such a time dependence is marginally consistent with the supernovae data \cite{Ellis:1998nz}, 
however at very late times the behaviour of the system needs to be modified.
The scale of this `quintessential' 
vacuum energy is the same as the scale of susy breaking on the brane, as can be seen from the explicit computation of the soft masses on the brane, 
hence asymptotically the observable brane becomes flat and supersymmetric. Of course, this is strictly true in the approximation
when one neglects the contribution of the matter sector. We have discussed
 the effective 4d supergravity that describes local dynamics of the brane-bulk system in the vicinity of the brane. It turns out that the 
supersymmetry breakdown in our model is seen on the brane as the kinetic breaking mediated to the observable sector by the radion multiplet, via  the 
matter-T modulus mixing in the K\"ahler function. 
To make the 
model suitable to describe realistically  
very late stages of the evolution of the Universe one needs to enhance the local scale of the brane supersymmetry breakdown with respect to the scale of the cosmological constant. The correspondence  of the 5d picture to 4d sugra with $T$-modulus suggests that the relevant enhancement mechanism 
would be equivalent to the tuning of the (super-)potential of the $T$-modulus and related to its stabilization.

\section{Appendix A: Global Structure of Space-time}

\subsection{Global Structure}
Let us  analyse the global structure of space-time for any value of $q$ and $C$. To do that we will define Kruskal 
coordinates covering space-time completely.
It is convenient to introduce tortoise coordinates
\begin{equation}
dr^*=f(r) dr,
\end{equation}
leading to
\begin{equation}
r^*=\frac{l^2}{2}  \frac{r_+}{r_+^2+r_-^2}  [\ln
(\frac{r-r_+}{r+r_+})+  \frac{r_-}{r_+}\arctan (\frac{r}{r_-})].
\end{equation}
We have defined $r_{\pm}$ as the roots of
\begin{equation}
r^4+ql^2r^2-Cl^2=(r^2+r_-^2)(r^2-r_+^2).
\end{equation}
More explicitly this leads to
\begin{equation}
r^2_{+}= \frac{-ql^2+\sqrt{l^4+C^2l^2}}{2}, \; r^2_{-}= \frac{ql^2+\sqrt{l^4+C^2l^2}}{2}.
\end{equation}
Notice that $r_+=l, \ r_-=0$ when $q=-1$ and $C=0$. 
The metric is originally defined for $r\ge r_+$. We will extend it to the origin.
Let us introduce the surface gravity
\begin{equation}
\kappa_{+}=\frac{r_+^2+r_-^2}{l^2r_+}
\end{equation}
Now the Eddington-Finkelstein coordinates,
\begin{equation}
u=t-r^*, v=t+r^*,
\end{equation}
correspond to null directions.
The Kruskal coordinates are 
\begin{equation}
U=-e^{-\kappa_+ u}, V=e^{\kappa_+ v}
\end{equation}
in such a way that
\begin{equation}
UV= -(\frac{r-r_+}{r+r_+})^2 e^{2\frac{r_-}{r_+}\arctan \frac{r}{r_-}}.
\end{equation}
The metric reads now
\begin{equation}
ds^2=-\frac{1}{\kappa_+^2 r^2}(r+r_+)^2 (r^2+r_-^2)e^{-2\frac{r_-}{r_+}\arctan \frac{r}{r_-}}dUdV
+r^2d\Sigma_q^2.
\end{equation}
As can be seen in the last expression the metric can be extended to $0<
r\le  r_+$ without any coordinate
singularity. After
doing so the product $UV$ goes from negative to positive values. The horizon
$r=r_+$ is mapped to the two lines $U=0$ and $V=0$. 
In particular one should notice that the regions on both sides of the 
Einstein-Rosen bridge at
$U=V=0$ are isometric and exchanged under the isometry $U\to -U, V\to -V$.

\subsection{Milne Branes}

Focusing on the case of branes receding away towards infinity, 
the motion is well approximated  by  the Friedmann equation 
\begin{equation} 
H^2=\frac{1}{r^2} 
\end{equation} 
where $H=\dot r /r$ with the 
proper time $\tau$ defined by 
\begin{equation} 
-(\frac{r^2}{l^2}-1)\dot t^2  +\frac{1}{\frac{r^2}{l^2}-1}\dot r^2=-1. 
\end{equation}  
This implies that
\begin{equation}
t=\frac{l}{2}\ln (\frac{\tau^2}{l^2}-1).
\end{equation}
This asymptotic situation corresponds to the case
$q=-1,\ C=0$. 
The geometry on the brane is dictated by the Milne metric
\begin{equation}
ds^2_B=-d\tau^2 +\tau^2d\Sigma^2_{-1}
\end{equation}
where
\begin{equation}
d\Sigma^2= d\chi^2+ \hbox{sinh}^2 \chi d\Omega^2
\end{equation}
and  $d\Omega^2$ is the metric on the unit two-sphere.
Locally  Milne space is isometric to Minkowski space upon 
using the following reparametrization
$
t=\tau\hbox{cosh}\chi,\ r=\tau \hbox{sinh}\chi,
$
leading to
\begin{equation}
ds^2_B=-dt^2+dr^2+r^2d\Omega^2.
\end{equation}
Therefore the brane is in fact a flat  
brane embedded in $AdS_5$.

\section{Appendix B: Brane Temperature}

Let us consider a free scalar field living on a Milne brane
\begin{equation}
ds^2=-dt^2+t^2d\Sigma^2
\end{equation}
The Klein Gordon equation can be easily reduced to a differential equation
by separation of variables
\begin{equation}
\Phi=J_k(\chi)\phi_k(t)
\end{equation}
where the $J_k$'s are   hyperbolic harmonics
\begin{equation}
\frac{\partial_{\chi}(\hbox{sinh}^2 \chi \partial_\chi J_k)}{\hbox{sinh}^2\chi}=-(k^2l^2+1)J_k
\end{equation}
and hyperbolic angular momenta $k$ are measured in units of $1/l$.
This leads to the Klein-Gordon equation
\begin{equation}
\ddot \phi_k+\frac{3}{t}\dot \phi_k +\frac{k^2l^2+1}{t^2}\phi_k=0
\end{equation} 
The modes are given by 
\begin{equation}
\phi_k^\pm=(\frac{t}{l})^{-1\pm ikl}.
\end{equation}
This defines the positive and negative frequency modes in Milne space.
One can quantize the fields by expanding
\begin{equation}
\Phi=\int dk (\phi_k^-a_k+ +\phi_k^- a_k^\dagger),
\end{equation}
where the operators $a_k$ and $a_k^\dagger$ are annihilation and creation operators
for the vacuum $\vert vac>_B$ on the brane.

The same quantization procedure for a static observer, $r=r_0=const.$, 
in the bulk whose position is
instantaneously coincident with the brane leads to 
the differential equation
\begin{equation}
\frac{d^2\phi_k}{dt_o^2}+\frac{k^2l^2+1}{l^2}\phi_k=0,
\end{equation} 
where
\begin{equation}
t_o=\frac{\sqrt{r^2_0-l^2}}{r}t.
\end{equation}
This leads 
 to the static observer's modes
\begin{equation}
\tilde \phi_k^\pm =e^{\pm i\Gamma \omega t},
\end{equation}
as in flat Minkowski space where
\begin{equation}
\omega=\frac{\sqrt{k^2l^2+1}}{l}
\end{equation}
and 
\begin{equation}
\Gamma= \frac{r_0}{\sqrt{r_0^2-l^2}}.
\end{equation}
We then identify $\Gamma \omega$ with the energy  of the particles with
hyperbolic angular momentum $k$. At high energy $\omega \gg 1/l $ and
large distances from the horizon $r_0\gg l$ we find that the energy
coincides with $k$ and $\omega$. 

We can then decompose fields as
\begin{equation}
\Phi=\int d\omega  (e^{-i\Gamma \omega t}\tilde
a_\omega +
+e^{i\Gamma \omega t}\tilde a_\omega^\dagger)
\end{equation}
where the operators $\tilde a_\omega $ and $\tilde a_\omega^\dagger$ are annihilation and creation operators
for the static bulk vacuum $\vert \tilde {vac}>$.

The two Hilbert spaces can be seen not to be unitarily equivalent.
Indeed one can decompose
\begin{equation}
\phi_k^=\int dp (A_{k\omega }e^{-i\Gamma\omega t}+ B_{k\omega
}e^{i\Gamma \omega t})
\end{equation}
from which we get
\begin{equation}
\tilde a_{\omega }=\int dk (A_{k\omega }a_k+ B^*_{k\omega }a^\dagger_k).
\end{equation}
The Bogoliubov coefficients depend on the overlap between the
eigenstates in the two Hilbert spaces
\begin{equation}
A_{k\omega}=\frac{\Gamma}{4\pi}\int \phi^-_ke^{i\omega t}dt.
\end{equation}
Similarly we have that
\begin{equation}
B_{k\omega}=A_{k,-\omega}.
\end{equation}
Using the explicit expression of $\phi^-_k$ one finds\cite{town}
\begin{equation}
B_{k\omega}=e^{-\pi kl}A_{k\omega}.
\end{equation}
Using the normalization conditions for  the Bogoliubov coefficients
\begin{equation}
(AA^{\dagger})_{kp}-(BB^{\dagger})_{kp}=\delta_{kp},
\end{equation}
we find that
\begin{equation}
(BB^{\dagger})_{kp}=\frac{\delta_{kp}}{e^{2\pi kl}-1}
\end{equation}
as a function of the hyperbolic angular momentum $k$. 
{}From this we conclude that the number of particles on the brane with
momentum $k$ 
is given by
\begin{equation}
N_{k}=\frac{1}{e^{2\pi l k}-1}
\end{equation}
i.e. a thermal spectrum with a temperature $1/2\pi l$.
 
It is more transparent to reexpress this result as a function of
the energy
\begin{equation}
E=\frac{dt}{d\tau} \Gamma \omega
\end{equation}
corresponding to the energy measured by an observer whose
clock ticks at the same rate as the clock on the moving brane.
This leads to 
\begin{equation}
N_{E}=\frac{1}{e^{2\pi \sqrt{E^2 (r^2-l^2) -1}}-1}
\end{equation}
corresponding to a thermal spectrum at high energy on the brane
with temperature
\begin{equation}
T_B=\frac{1}{2\pi l\sqrt{\frac{r^2}{l^2}-1}}.
\end{equation}

At large distances, when the brane is far away from the horizon we
find that the temperature is
\begin{equation}
T_B=\frac{1}{2\pi \tau},
\end{equation}
where $\tau$ is the proper time on the brane.

\vspace{1.5cm} 
%\noindent {\bf Acknowledgments}   
\noindent We thank John Ellis and Patrizia Bucci for very interesting 
discussions.   
Z.L. thanks Theory Division at CERN for hospitality.
\noindent This work  was partially supported  by the EC Contracts
HPRN-CT-2000-00152 and HPRN-CT-2000-00148 for years 2000-2004, by the Polish State Committee for Scientific Research grants, and by POLONIUM 2003.

\end{document}